\documentclass[conference]{IEEEtran}
\IEEEoverridecommandlockouts
\usepackage{cite}
\usepackage[american]{circuitikz}
\usepackage{tikz} 
\usepackage{multirow}
\usepackage{xcolor}
\usepackage{tikz}
\usepackage{array}

\usepackage[percent]{overpic}
\usepackage{amsmath,amssymb,amsfonts}
\usepackage{algorithmic}
\usepackage{graphicx}
\usepackage{soul}
\usepackage{comment}
\usepackage{geometry}
\usetikzlibrary{positioning,arrows.meta}

\usepackage{textcomp}
\usepackage{xcolor}
\def\BibTeX{{\rm B\kern-.05em{\sc i\kern-.025em b}\kern-.08em
    T\kern-.1667em\lower.7ex\hbox{E}\kern-.125emX}}
\begin{document}

\title{A comparative study on power delivery aspects of compute-in/near-memory approaches using DRAM}


\author{
Siddhartha Raman Sundara Raman, Siyuan Ma, Lizy Kurian John \\
The Department of Electrical and Computer Engineering, The University of Texas at Austin, Austin, Texas, 78712 \\
s.siddhartharaman@utexas.edu
}

\maketitle

\begin{abstract}

Compute-in-memory (PIM) mitigates the memory wall by performing computation within memory, reducing data movement and improving energy efficiency. DRAM-based PIM is particularly attractive due to its high density, mature manufacturing ecosystem, and compatibility with existing systems. Recent works exploit multiple levels of the DRAM hierarchy—including subarrays, banks, and 3D-stacked organizations—to enable in-memory computation using mechanisms such as multi-row activation, row-buffer operations, and near-bank compute units. However, these approaches introduce non-traditional current demand patterns that challenge the power delivery network (PDN).

This paper surveys PDN challenges in DRAM-based PIM systems and proposes a unified taxonomy that characterizes PIM-induced current behavior along temporal (burst vs. sustained) and spatial (localized vs. distributed) dimensions. Using this framework, we analyze how representative PIM techniques stress the PDN through bursty activations, multi-row concurrency, and large-scale parallel execution, leading to voltage droop, IR drop, and thermal hotspots.

We further discuss DRAM-specific mitigation strategies leveraging existing architectural and circuit-level mechanisms, including timing constraints, memory controller scheduling, data placement, and bank- and vault-level power management. This survey highlights the importance of PDN-aware design for scalable and reliable DRAM-based PIM systems and outlines key future research directions.

\end{abstract}

\section{Introduction}
The continued scaling of data-intensive applications, including machine learning, scientific computing, and large-scale analytics, has intensified the long-standing “memory wall” problem in conventional Von Neumann architectures. The separation between compute and memory leads to excessive data movement, which dominates both system latency and energy consumption. As a result, improving overall system efficiency increasingly requires reducing data transfer and exploiting memory-centric computation.

Compute-in-memory (CIM), or processing-in-memory (PIM), has emerged as a promising paradigm to address this challenge by enabling computation within or near memory arrays. A wide range of CIM approaches has been proposed, spanning digital and analog execution models, near-memory accelerators, and fully in-situ computation. These approaches have been explored across diverse memory technologies, including SRAM \cite{eChimera}\cite{GCN}\cite{Conv_SRAM}\cite{Shanshan_Ising}\cite{ABI}\cite{SACHI}\cite{Ising_arxiv}\cite{NEM_GNN_arxiv}\cite{SPARK_arxiv}\cite{Cryo_arxiv}\cite{SPARK}, embedded DRAM (eDRAM) \cite{eDRAM_1} \cite{IGZO_CIM}\cite{UT_Thesis}, and emerging non-volatile memories  such as resistive RAM (RRAM)\cite{RRAM_1}\cite{RRAM_2}\cite{RRAM_3}\cite{RRAM_4}\cite{RRAM_cache} phase-change memory \cite{NVM_Raman}(PCM)\cite{PCM_1}\cite{PCM_2}, and phase transition material (PTM)\cite{8T_SRAM_1} \cite{6T_SRAM} enabled designs, magneto-resistive Random Access Memories \cite{MRAM_1}\cite{MRAM_2}, Ferroelectric field effect transistors (FeFET) enabled memories \cite{FeFET}\cite{FeFET_CIM}. While these technologies offer different trade-offs in density, latency, and energy efficiency, commodity DRAM \cite{ComputeDRAM} \cite{DRISA}\cite{AMBIT} remains particularly attractive due to its high density, low cost, mature ecosystem, and widespread deployment in modern computing systems for large-scale storage.

Recent works that have focused on DRAM-based CIM, leverage the internal organization of DRAM banks and subarrays to perform computation using existing structures such as sense amplifiers, bitlines, and row buffers. Techniques such as bulk bitwise operations, row-clone mechanisms, and subarray-level parallelism \cite{SALP}\cite{DRAM_perf} enable data-parallel computation within DRAM while maintaining compatibility with standard memory interfaces. These approaches exploit the inherent structure of DRAM to achieve significant improvements in throughput and efficiency, without requiring fundamentally new memory technologies.

However, a critical and often underexplored aspect of DRAM-based CIM is the power delivery network (PDN). Conventional DRAM systems are designed under the assumption of relatively constrained and well-regulated access patterns, enforced through timing parameters such as activation spacing and current limits. In contrast, DRAM-based CIM architectures introduce significantly higher levels of concurrency, including overlapping row activations, multi-subarray activity, and bulk data operations. These changes lead to bursty and spatially correlated current demand, which can stress the PDN and result in sudden voltage droop\cite{DRAM_power}\cite{UTBB_SOI}\cite{JJFET}, IR drop\cite{IR_drop}\cite{Ising_summary}, and thermal hotspots.

The challenge is further amplified by the hierarchical organization of DRAM. At the bank level, multiple subarrays share global resources such as bitlines and peripheral circuitry, while across banks and ranks, simultaneous operations can create large aggregate current demands at the channel and module level. Although timing constraints such as tRRD and tFAW are designed to limit activation-induced current surges, emerging CIM techniques may push these constraints to their limits or operate in regimes not originally anticipated by standard DRAM specifications. As a result, understanding and managing PDN becomes critical to ensuring both performance and correctness.

Despite its importance, PDN-aware design in DRAM-based CIM remains insufficiently characterized. Many prior works focus primarily on architectural performance gains and assume idealized or simplified power models, without fully accounting for realistic current limits, activation windows, or voltage fluctuations. This gap makes it difficult to assess the scalability and feasibility of proposed designs, particularly those that rely on aggressive forms of intra/inter-bank parallelism.

In this survey, we present a comprehensive study of power delivery challenges in DRAM-based CIM systems. We first provide an overview of DRAM-centric compute mechanisms and the forms of parallelism they exploit, including bank-level,  and subarray-level parallelism. We then analyze how these mechanisms impact PDN behavior across different levels of the memory hierarchy, from subarray and bank-level current draw to 3D-level power constraints. Finally, we review existing techniques for mitigating PDN challenges, including scheduling techniques, current limiting, and architectural adaptations, and outline key directions for future research in PDN-aware DRAM-based computing
\begin{figure}[t]
\centering
\resizebox{0.95\columnwidth}{!}{%
\begin{tikzpicture}[
    font=\footnotesize,
    every node/.style={
        draw,
        rectangle,
        minimum height=0.5cm,
        align=center,
        inner sep=2pt
    },
    edge from parent/.style={draw, -latex, thick},
    level distance=1.0cm,
    level 1/.style={sibling distance=3.0cm},
    level 2/.style={sibling distance=1.8cm},
    level 3/.style={sibling distance=1.6cm},
    level 4/.style={sibling distance=2.2cm}
]

\node[minimum width=1.9cm, fill=gray!20] {DRAM System}
    child {node[minimum width=1.45cm, fill=blue!15] {Channel 0}}
    child {node[minimum width=1.2cm, fill=gray!10] {$\cdots$}
        child {node[minimum width=1.2cm, fill=green!20] {Rank 0}
            child {node[minimum width=1.1cm, fill=orange!20] {Bank 0}
                child {node[minimum width=1.35cm, fill=red!15] {Subarray 0}}
                child {node[minimum width=1.35cm, fill=red!15] {Subarray 1}}
            }
            child {node[minimum width=1.1cm, fill=orange!20] {Bank 1}}
        }
        child {node[minimum width=1.2cm, fill=green!20] {Rank 1}}
    }
    child {node[minimum width=1.45cm, fill=blue!15] {Channel N}};

\end{tikzpicture}%
}
\caption{DRAM hierarchy with color-coded levels: channels (blue), ranks (green), banks (orange), and subarrays (red).}
\label{fig:dram_hierarchy_centered}
\end{figure}
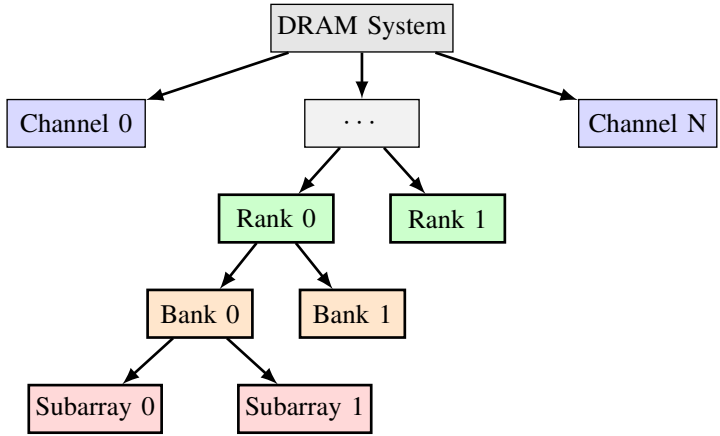

\section{PIM using DRAM}
\subsection{Organization/bitcell}
Dynamic Random-Access Memory (DRAM) is the predominant main memory technology in modern computing systems due to its high storage density and cost efficiency. A DRAM system is hierarchically organized to balance performance, scalability, and bandwidth. At the highest level, memory is divided into channels, each providing an independent interface between the memory controller and DRAM modules. Within each channel, memory is further partitioned into ranks, where a rank consists of multiple DRAM chips that operate in lockstep and share command and address signals. Each rank is subdivided into multiple banks, which can operate largely independently, enabling parallelism at the bank level, shown in Fig.\ref{fig:dram_hierarchy_centered}. Internally, each bank is composed of multiple subarrays, forming a two-dimensional organization of rows and columns that store the actual data.

\begin{figure}[t]
\centering

\begin{minipage}[t]{0.40\columnwidth}
\centering
\vspace{0pt}

\makebox[0pt][l]{(a)}\vspace{-0.6em}

\begin{tikzpicture}[line width=0.9pt, scale=0.78, every node/.style={font=\scriptsize}]

\draw (0,4.2) -- (0,0.2);
\node[above] at (0,4.2) {\textbf{BL}};

\draw (-1.2,3.4) -- (3.6,3.4);
\node[left] at (-1.2,3.4) {\textbf{WL}};

\draw (1.6,3.4) -- (1.6,2.7);
\draw (1.25,2.7) -- (1.95,2.7);

\draw (1.15,2.5) -- (2.05,2.5);
\draw (1.15,2.5) -- (1.15,1.7);
\draw (2.05,2.5) -- (2.05,1.7);

\draw (0,1.7) -- (1.15,1.7);

\draw (2.05,1.7) -- (3.0,1.7);

\draw (3.0,1.7) -- (3.0,1.25);
\draw (2.7,1.25) -- (3.3,1.25);
\draw (2.7,1.05) -- (3.3,1.05);
\draw (3.0,1.05) -- (3.0,0.65);

\draw (2.75,0.65) -- (3.25,0.65);
\draw (2.75,0.65) -- (3.0,0.3);
\draw (3.25,0.65) -- (3.0,0.3);

\end{tikzpicture}
\end{minipage}
\hfill
\begin{minipage}[t]{0.56\columnwidth}
\centering
\vspace{0pt}

\makebox[0pt][l]{(b)}\vspace{0.3em}

\scriptsize
\renewcommand{\arraystretch}{1.12}
\setlength{\tabcolsep}{4pt}

\begin{tabular}{|>{\centering\arraybackslash}m{1.15cm}|>{\raggedright\arraybackslash}m{3.0cm}|}
\hline
\textbf{Parameter} & \textbf{Description} \\
\hline
$t_{RCD}$   & ACT to READ/WRITE delay \\
\hline
$t_{RP}$    & PRECHARGE latency \\
\hline
$t_{RAS}$   & Minimum row active time \\
\hline
$t_{RC}$    & Row cycle time \\
\hline
$t_{RRD}$   & ACT to ACT delay \\
\hline
$t_{FAW}$   & Four ACT window \\
\hline
$t_{CCD}$   & Column-to-column delay \\
\hline
$t_{BURST}$ & Data burst duration \\
\hline
\end{tabular}
\end{minipage}

\caption{(a) 1T1C DRAM bitcell (b) Representative DRAM timing parameters.}
\label{fig:dram_bitcell_timing}
\end{figure}
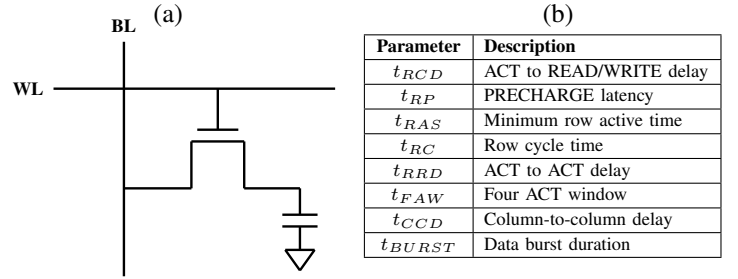

At the circuit level, each DRAM cell consists of a capacitor and an access transistor. The capacitor stores data as charge, while the transistor controls access to the cell, shown in Fig.\ref{fig:dram_bitcell_timing}a). Due to charge leakage over time, DRAM cells require periodic refresh operations to maintain data integrity. Within each subarray, rows of cells are connected to local bitlines and interfaced with sense amplifiers, which collectively form the row buffer. When a row is accessed, the entire row is activated and its contents are sensed and latched into the row buffer. Subsequent accesses to the same row can be served directly from the row buffer without re-accessing the cell array, resulting in lower latency compared to accessing a different row.
DRAM operations are governed by a sequence of commands that control data movement between the cell array, row buffer, and external interface. An ACTIVATE command opens a row by transferring its contents into the row buffer. Once the row is active, READ or WRITE commands access specific columns of data via global bitlines and the memory channel. Before a different row within the same bank can be accessed, a PRECHARGE command is issued to close the currently active row and restore the bitlines to a neutral state. These operations are constrained by timing parameters such as the ACTIVATE-to-READ delay (tRCD), row precharge latency (tRP), minimum row active time (tRAS), and row cycle time (tRC), as well as constraints on successive activations (e.g., tRRD and tFAW), which collectively ensure correct operation and regulate power consumption, shown in Fig.\ref{fig:dram_bitcell_timing}b).

\subsection{Degrees of parallelism}
The hierarchical organization of DRAM naturally exposes multiple degrees of parallelism across different levels of the memory system. At the coarsest granularity, channels operate independently, enabling parallel data transfers across multiple memory interfaces. Within each channel, ranks provide an additional level of parallelism, where operations such as activation and precharge can be overlapped across ranks, even though only one rank drives the data bus at a time. At a finer granularity, banks within a rank can be accessed largely independently, allowing concurrent memory operations and improving overall throughput. Further, within each bank, subarrays offer an additional degree of parallelism, as they contain separate local sensing circuitry and can support partially overlapping operations depending on timing constraints. However, this parallelism is not uniform across all operations, as shared global resources such as bitlines and I/O interfaces introduce serialization points, particularly for data transfer. Understanding these different levels of parallelism is critical for analyzing both performance scaling and power delivery behavior in DRAM systems.
\subsection{Different levels of PIM using DRAM}
Broadly speaking, there are 3 levels of processing in-DRAM techniques that have been proposed namely 2D sub-array level PIM, 2D bank-level PIM, and 3D PIMs, that involve compute using memory arrays that involve vertical stacking.

\begin{table}[t]
\centering
\caption{Summary of representative PIM approaches across different levels of the DRAM hierarchy.}
\label{tab:pim_summary}
\scriptsize
\renewcommand{\arraystretch}{1.15}
\setlength{\tabcolsep}{4pt}

\resizebox{\columnwidth}{!}{%
\begin{tabular}{
|>{\centering\arraybackslash}p{2.8cm}
|>{\centering\arraybackslash}p{1.2cm}
|>{\centering\arraybackslash}p{4.2cm}|
}

\hline
\textbf{Category} & \textbf{Example} & \textbf{Method} \\
\hline

\multirow{3}{=}{Sub-array level PIM}
& AMBIT
& 3-row activation for AND/OR, separate cells for NOT \\
\cline{2-3}
& DRISA
& Near-sense amplifier logic for computing NOR, addition \\
\cline{2-3}
& RowClone
& Row buffer for fast copy \\
\hline

\multirow{2}{=}{Bank-level PIM}
& Newton
& Bank periphery addition of adder tree, multipliers \\
\cline{2-3}
& HBM-PIM
& Bank periphery addition of registers, control, adder, multiplier \\
\hline

\multirow{3}{=}{3D level PIM\\(using Hybrid Memory Cube)}
& Neurocube
& Each vault has programmable neurosequence generator and processing element \\
\cline{2-3}
& Tetris
& Adds global buffer for input data reuse to Neurocube-like structure \\
\cline{2-3}
& iPIM
& Decouples control from execution units \\
\hline

\end{tabular}%
}
\end{table}
\subsubsection{Sub-array level PIM}
Subarray-level processing-in-memory (PIM) techniques exploit the internal organization of DRAM banks, particularly the presence of subarrays with local sense amplifiers and row buffers, to enable computation directly within the memory array. A key direction in this space is cell-level PIM, which integrates or leverages logic at the bitline sense amplifiers to execute bulk operations across entire rows, thereby maximizing internal memory bandwidth. However, directly embedding logic within DRAM cells is challenging due to the extremely tight cell pitch, which is optimized for only a single transistor and capacitor. As a result, most prior works avoid explicit logic gate integration and instead exploit the analog behavior of DRAM circuitry to realize simple operations. For example, AMBIT\cite{AMBIT} introduces triple-row activation (TRA), where three wordlines are activated simultaneously, causing charge sharing on the bitline. The sense amplifier resolves the resulting value based on a majority function, enabling bulk bitwise operations such as AND and OR by appropriately initializing one of the rows. Ambit further proposes dual-contact cells (DCCs) to support NOT operations by transferring inverted sense amplifier values back to the cell, although such modifications introduce layout challenges. In contrast, DRISA\cite{DRISA} explores augmenting DRAM with simple logic structures near the sense amplifiers, proposing modified cell organizations (e.g., 3T1C-NOR and 1T1C-based designs) and additional circuitry to enable operations such as NOR, addition, and data movement across bitlines. While these approaches increase functionality, integrating additional transistors within the DRAM pitch remains a significant challenge, and some designs rely on alternative cell structures or peripheral logic to maintain feasibility. Complementing these logic-centric approaches, RowClone \cite{RowClone} focuses on efficient data movement by leveraging the row buffer to perform fast bulk copy operations within a subarray using back-to-back activations (Fast Parallel Mode). Notably, AMBIT and DRISA build upon such mechanisms to accelerate data preparation and movement, and introduce optimizations such as multi-row activation support and fused command sequences (e.g., activate–activate–precharge) to reduce latency. Collectively, these subarray-level PIM techniques demonstrate how DRAM’s intrinsic structures can be repurposed to perform compute with minimal modifications, while also highlighting practical challenges related to device scaling, control complexity, and power delivery, which will be discussed later in this article.
\subsubsection{Bank-level PIM}
Beyond subarray-level techniques, bank-level PIM architectures have been proposed to address the practical challenges associated with cell-level integration, particularly the severe area constraints imposed by the shrinking pitch of DRAM cells. While cell-level PIM maximizes internal bandwidth by operating directly on entire rows, its feasibility is limited due to the difficulty of embedding additional logic within tightly optimized DRAM cells. To overcome this limitation, approaches such as Newton and HBM-PIM integrate compute logic at the periphery of the bank, typically after the column decoder and selector of the banks, thereby allowing computation to leverage the full width of the cell array without modifying the cell structure. These designs trade off fine-grained, row-level parallelism for coarse-grained bank-level parallelism, where multiple banks operate concurrently to compensate for reduced per-bank internal bandwidth utilization.

In particular, Newton\cite{Newton} targets memory-bound deep learning workloads by embedding fixed-function compute units within each bank, consisting of multipliers, adder trees, and accumulation registers to efficiently perform matrix-vector multiplication. It introduces specialized PIM commands (e.g., global write, multi-bank activation, compute, and result readback) to orchestrate data movement and computation across banks, enabling simultaneous execution of multiply-accumulate (MAC) operations. Similarly, HBM-PIM \cite{HBM_PIM} integrates programmable compute units (PCUs) shared across banks, supporting SIMD-style arithmetic operations and enabling more flexible execution models. These designs exploit bank-level parallelism, high-bandwidth memory organization, along with integration of near-bank logic, to scale performance, while maintaining compatibility with DRAM manufacturing constraints.

\subsubsection{3D PIM}
Beyond bank-level PIM, 3D-stacked PIM architectures extend compute capabilities by leveraging both the logic die and vertically stacked memory dies, enabling tighter integration and higher internal bandwidth. Unlike bank-level PIM approaches such as Newton and HBM-PIM, where compute logic is placed near the periphery of memory banks within a single die, 3D PIM architectures utilize a dedicated logic layer beneath stacked memory layers to enable more energy-efficient communication through through-silicon vias (TSVs). Architectures such as Neurocube \cite{Neurocube}, Tetris \cite{Tetris}, and iPIM \cite{iPIM} explore this paradigm using Hybrid Memory Cube (HMC), which partitions memory into multiple vertical vaults. In Neurocube, each vault is paired with a programmable neurosequence generator and processing elements (PEs) connected via an on-chip network to enable parallel neural network computation. Tetris extends this design by incorporating per-vault global buffers to improve data reuse efficiency, while iPIM further enhances parallelism by decoupling control and execution across logic and memory dies, and introducing a single-instruction-multiple-bank execution model to exploit bank-level bandwidth within each vault.

\begin{table}[t]
\centering
\renewcommand{\arraystretch}{1.5} 
\setlength{\tabcolsep}{10pt} 
\caption{Unified PDN taxonomy for DRAM-based PIM techniques categorized by temporal and spatial characteristics.}
\label{tab:pdn_taxonomy}
\begin{tabular}{|c|c|c|}
\hline
 & \textbf{Localized} & \textbf{Distributed} \\
\hline
\textbf{Burst} 
& RowClone 
& AMBIT \\
\hline
\textbf{Sustained} 
& DRISA 
& Newton, HBM-PIM, Neurocube, Tetris, iPIM \\
\hline
\end{tabular}
\end{table}

\section{Power delivery network aspects}
\subsection{Problems}
The impact of compute-in-DRAM techniques on the power delivery network (PDN) can be systematically understood using a unified taxonomy based on the temporal characteristics (burst vs. sustained) and spatial distribution (localized vs. distributed) of current demand. This framework provides a consistent lens for analyzing diverse PIM architectures spanning subarray-level, bank-level, and 3D designs, as shown in Table II.

At the subarray level, early PIM mechanisms such as RowClone and Ambit are primarily characterized by burst-dominated current behavior. RowClone induces short-duration, spatially localized current spikes due to back-to-back ACTIVATE operations within a subarray, resulting in high di/dt stress on local PDN structures such as wordline drivers and sense amplifiers. In contrast, Ambit introduces bursty but spatially correlated current demand, where simultaneous multi-row activation along shared bitlines leads to significantly higher instantaneous current density and increased sensitivity to voltage droop during sense amplification. More advanced subarray-level designs such as DRISA extend this behavior into the sustained regime, where repeated logic operations and in-memory computation result in prolonged and spatially distributed current draw across multiple subarrays within a bank. Consequently, subarray-level PIM spans a spectrum from localized burst events to moderately distributed sustained activity, primarily stressing intra-bank PDN components.

At the bank level, architectures such as Newton and HBM-PIM expand both the spatial and temporal scope of PDN stress. By exploiting bank-level parallelism, these designs enable concurrent activation and computation across multiple banks, resulting in spatially distributed current demand at the rank level. In the unified taxonomy, bank-level PIM exhibits mixed temporal behavior, combining burst components from coordinated activation phases with sustained current draw during parallel compute execution (e.g., MAC operations). This shifts the PDN bottleneck from localized subarray structures to shared global supply networks, including bank periphery, global bitlines, and rank-level power rails, increasing the likelihood of aggregate voltage droop and supply noise.

At the 3D-stacked level, architectures such as Neurocube, Tetris, and iPIM further extend this taxonomy into the fully distributed and sustained regime across both horizontal and vertical dimensions. These systems leverage vault-level parallelism and logic-memory integration through TSVs, resulting in concurrent activity across multiple banks, vaults, and stacked dies. As a result, the PDN must support highly correlated current demand spanning multiple layers, introducing additional challenges such as TSV-induced IR drop, inter-die voltage coupling, and thermal hotspots. In particular, hybrid designs such as iPIM, which combine bank-level parallelism with 3D integration, represent the extreme case of the taxonomy, where both temporal concurrency and spatial distribution are maximized, leading to complex multi-scale PDN interactions.
\subsection{Mitigation strategies}
Mitigating PDN challenges in DRAM-based PIM systems requires leveraging DRAM-specific timing and architectural mechanisms that inherently regulate current demand \cite{DRAM_power}\cite{DRAM_power_2}\cite{DRAM_power_3}. Commodity DRAM systems enforce activation constraints such as tRRD (row-to-row delay) and tFAW (four activate window) to limit the number of concurrent row activations and bound peak current draw at the bank and rank level \cite{jedec2012ddr4}. These constraints can be extended or made adaptive to regulate bursty current behavior introduced by PIM operations, as activation-driven current dominates instantaneous power consumption in DRAM \cite{ghose2018vampire, vogelsang2010energy}. In addition, memory controller scheduling policies can be used to distribute ACTIVATE commands across banks and ranks, avoiding synchronized current surges while maintaining high throughput \cite{rixner2000memory, mutlu2008parallelism}.
At the subarray level, where PIM mechanisms introduce elevated instantaneous current (e.g., multi-row activation), mitigation can be achieved through subarray-aware scheduling and data placement, ensuring that high-current operations are spatially separated within a bank \cite{arxiv2024subarray}. DRAM designs also incorporate on-die decoupling capacitance near sense amplifiers and wordline drivers to stabilize supply voltage during activation and sensing; enhancing or strategically placing these decoupling capacitors can help absorb transient current spikes caused by burst-dominated PIM operations. Since activation and sensing phases dominate current spikes, localized mitigation at the subarray level is particularly important \cite{vogelsang2010energy}.
For sustained and spatially distributed workloads, such as those arising in bank-level PIM, mitigation relies on bank-aware scheduling and activation spreading, which exploits bank-level parallelism while avoiding simultaneous peak demand across multiple banks \cite{mutlu2008parallelism}. DRAM systems allow overlapping internal operations across banks while serializing data transfer, enabling the memory controller to shape current demand temporally \cite{rixner2000memory}. Such scheduling techniques are critical for preventing aggregate current surges at the rank level, where multiple banks share global power delivery resources.
In 3D-stacked DRAM systems, such as HBM and HMC, mitigation extends to vault-level load balancing and TSV-aware power delivery design \cite{gao2021survey}. The partitioning of memory into vaults provides a natural granularity for distributing activity and avoiding localized current concentration. Dedicated power TSVs and hierarchical power distribution networks across stacked dies can be optimized to reduce IR drop and current crowding under high parallel activity. Furthermore, the strong coupling between power and temperature in stacked memory systems necessitates thermal-aware scheduling, as increased current density directly impacts reliability and performance \cite{pavlidis2023thermal}.
Overall, these DRAM-centric mitigation strategies—ranging from timing constraint enforcement and controller-level scheduling to subarray-aware mapping and vault-level load balancing—demonstrate that PDN robustness in PIM-enabled DRAM systems can be achieved by carefully orchestrating existing DRAM mechanisms to manage both burst and sustained current demand \cite{ghose2018vampire, gao2021survey}.

\section{Conclusion}

DRAM-based compute-in-memory (PIM) offers a promising path to overcome the memory wall by reducing data movement and exploiting memory-level parallelism. However, these approaches fundamentally alter current demand patterns, introducing new challenges for the power delivery network (PDN). In this work, we presented a unified taxonomy based on temporal (burst vs. sustained) and spatial (localized vs. distributed) characteristics to systematically analyze PDN behavior across subarray-, bank-, and 3D-stacked PIM architectures.

Our analysis shows a clear progression: subarray-level techniques induce localized, burst-dominated current spikes; bank-level designs introduce spatially distributed and partially sustained demand; and 3D-stacked systems extend these effects into fully distributed, vertically coupled current profiles. To address these challenges, we discussed DRAM-specific mitigation strategies, including timing constraint enforcement, scheduling across banks and subarrays, and vault-level load balancing. Overall, enabling scalable DRAM-based PIM requires PDN-aware design across the memory hierarchy, highlighting the need for tighter integration between architecture, system control, and power delivery mechanisms in future memory systems.

\bibliographystyle{IEEEtran}
\bibliography{refs}

\end{document}